\documentclass{PoS}
\pdfoutput=1
\usepackage{amsmath,amssymb,amsbsy,multicol,url}
\usepackage[utf8]{inputenc}
\usepackage[numbers,sort&compress]{natbib}
\usepackage{natbibspacing}
\usepackage{graphicx}
\usepackage{ulem,color}
\newcommand{\TXTheight}{\textheight}

\title{Large scale separation and hadronic resonances from a new strongly interacting sector}

\ShortTitle{Large scale separation and hadronic resonances}

\author{Anna Hasenfratz \\
  Department of Physics, University of Colorado, Boulder, CO, USA\\
  E-mail: \email{Anna.Hasenfratz@colorado.edu}}

\author{Claudio Rebbi\\
  Department of Physics, Boston University, Boston, MA, USA \\
  E-mail: \email{rebbi@bu.edu}}

\author{\speaker{Oliver Witzel}\thanks{Present address: Department of Physics, University of Colorado, Boulder, CO, USA} \\
  Higgs Centre for Theoretical Physics, School of Physics \& Astronomy,\\
  The University of Edinburgh, Edinburgh, United Kingdom \\
  E-mail: \email{Oliver.Witzel@colorado.edu}}

\abstract{A large separation of scales is frequently required by theories describing physics beyond the Standard Model. In mass-split models with  some massless (light) and some heavy flavors, large scale separation arises  by construction if the system is conformal in the ultraviolet but chirally broken in the infrared. Due to the presence of a conformal fixed point, such chirally broken systems show hyperscaling and have a highly constrained resonance spectrum that is significantly different from the QCD spectrum. We present numerical evidence for hyperscaling of both  light-light and heavy-heavy meson resonances and show that they only depend on the ratio of the light and heavy flavor masses.  The heavy-heavy spectrum is qualitatively different from QCD. The mass of heavy-heavy quarkonia e.g.~is not proportional to the constituent quark mass. }

\FullConference{The European Physical Society Conference on High Energy Physics\\
		5-12 July, 2017\\
		Venice}

\begin{document}

\section{Introduction}
With the discovery of the Higgs boson in 2012, an essential piece has been added to the Standard Model (SM) but important questions regarding the UV completeness of the SM or the nature of electro-weak symmetry breaking remain open. Experimentally the Large Hadron Collider (LHC) is exploring the low TeV range but so far no direct signals of new physics have been discovered. Thus a theory describing electro-weak symmetry breaking beyond the Standard Model (BSM) has to accommodate a 125 GeV Higgs boson, while other states must me much heavier (at least a few TeV) to have escaped  experimental searches. We are therefore seeking theories exhibiting a large separation of scales but at the same time satisfying electro-weak precision constraints.  This means any viable model must have properties quite different from quantum chromodynamics (QCD).

\begin{figure}[!h]
  \centering
  \begin{picture}(125,20)
    \put(0,10){UV}    \put(6,11){\thicklines{\vector(1,0){110}}} \put(120,10){IR}
    \put(8,7){$\Lambda_{UV}$} \put(63,7){$\Lambda_{IR}$}
    \put(29,13){conformal} \put(25,7){fermion masses}
    \put(80,13){chirally broken} \put(80,7){Higgs dynamics}
  \end{picture}
  \caption{Sketch motivating the idea of mass-split systems. In our case the chirally broken system is given by an SU(3) gauge theory with four light (massless) flavors and we add eight heavy flavors  such that our system feels the attraction of the conformal IRFP of  degenerate 12 flavors.}
  \label{Fig.ScaleSep}  
\end{figure}
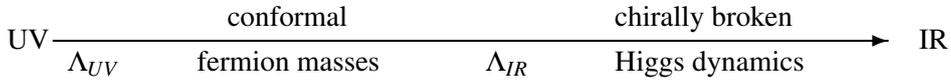

One possible scenario  for the Higgs boson is to be a composite particle described by a new strongly interacting  gauge-fermion system. Possible models for such a scenario are chirally broken in the IR but conformal in the UV and  automatically lead to a large separation of scales, see e.g.~\cite{Luty:2004ye,Dietrich:2006cm,Vecchi:2015fma,Ferretti:2013kya}. 
The general idea is sketched in Fig.~\ref{Fig.ScaleSep} where we start with a chirally broken massless theory in the IR on the right. Next we add additional, massive flavors to push the system into the conformal regime at higher energies. These massive flavors decouple in the IR and the Higgs boson emerges from the massless flavors of the new strongly interacting sector. The masses of fermions and gauge bosons arise from interaction terms of the new sector with the SM fields. This model gives an effective description of the electro-weak sector but requires itself a UV completion to explain the origin of the heavy masses.

Here we report on one such model \cite{Brower:2014dfa,Brower:2015owo,Hasenfratz:2016gut} based on an SU(3) gauge theory with fundamental flavors and focusing only at the new strongly interacting sector in isolation. In the IR we have four degenerate flavors (considered to be massless) and add eight heavy flavors to push the system near the infrared fixed-point (IRFP) of the twelve flavor conformal theory.
This model allows for two possibilities for the Higgs boson to emerge: 1)  as the iso-singlet scalar ($0^{++}$) of the massless flavors, a dilaton-like particle arising from the broken conformal symmetry; 2) as a pseudo Nambu-Goldstone boson similar to pions in QCD. In the latter case, it can e.g.~be embedded in a 2-Higgs doublet type model adjusting its quantum numbers \cite{Ma:2015gra}. 

In the next Section we present some details of our model with four light and eight heavy flavors and show highlights of our nonperturbative investigations. We emphasize the key features  making our model highly predictive and distinct from QCD, also focusing at the iso-singlet scalar ($0^{++}$) state. Some concluding remarks and a brief outlook will follow.

\section{The four plus eight model and the iso-singlet scalar}
Using lattice field theory techniques, we perform nonperturbative investigations of an SU(3) gauge theory with four light flavors of mass $m_\ell$ and eight heavy flavor with mass $m_h$. The choice of 4, 8, and 12 flavors is motivated by performing exploratory simulations with numerically inexpensive staggered fermions; in the future simulations e.g.~with domain-wall fermions are however warranted \cite{Hasenfratz:2017mdh}.

Our model exhibits two key properties: it breaks chiral symmetry but at the same time it also shows hyperscaling like it is expected for conformal field theories. Wilsonian renormalization group arguments imply that systems near a conformal fixed-point exhibit hyperscaling i.e.~dimensionless ratios of hadron masses (or amplitudes) are independent of $m_h$ both in the $m_\ell=m_h$ mass-degenerate and the $m_\ell=0$ chiral limits \cite{DelDebbio:2010ze,Brower:2015owo}. We generalized these  relations for arbitrary masses in Ref.~\cite{Hasenfratz:2016gut} and showed that dimensionless ratios of hadron masses (or amplitudes)  depend only on the ratio of the flavor masses $m_\ell/m_h$, a property that makes  mass-split models  highly predictive. Thus we present our numerical data in terms of the dimensionless ratio $m_\ell/m_h$.

\begin{figure}[tb]
  \centering
  \parbox{0.49\textwidth}{\includegraphics[height=0.25\textheight]{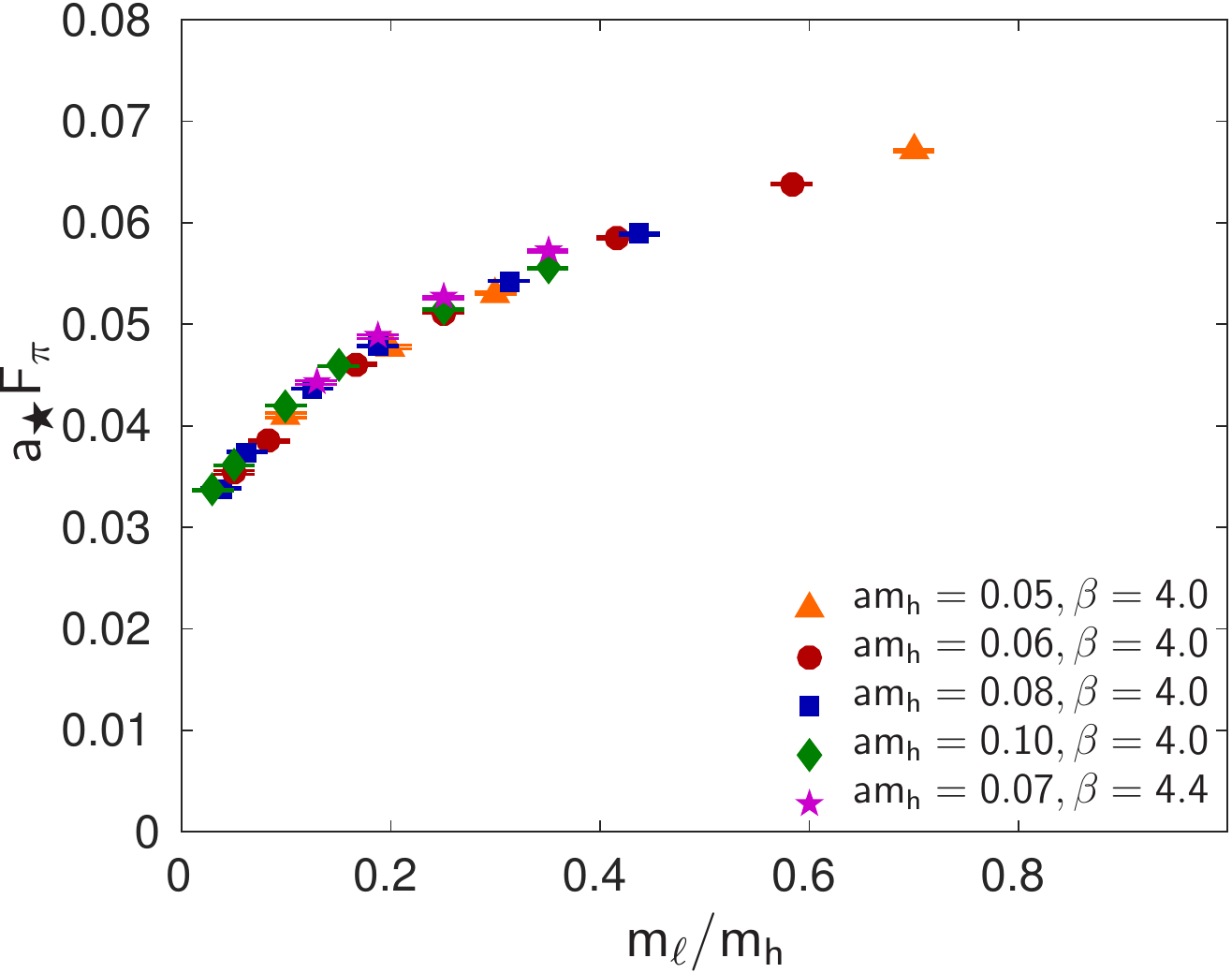}}\hfill
  \parbox{0.49\textwidth}{\includegraphics[height=0.25\textheight]{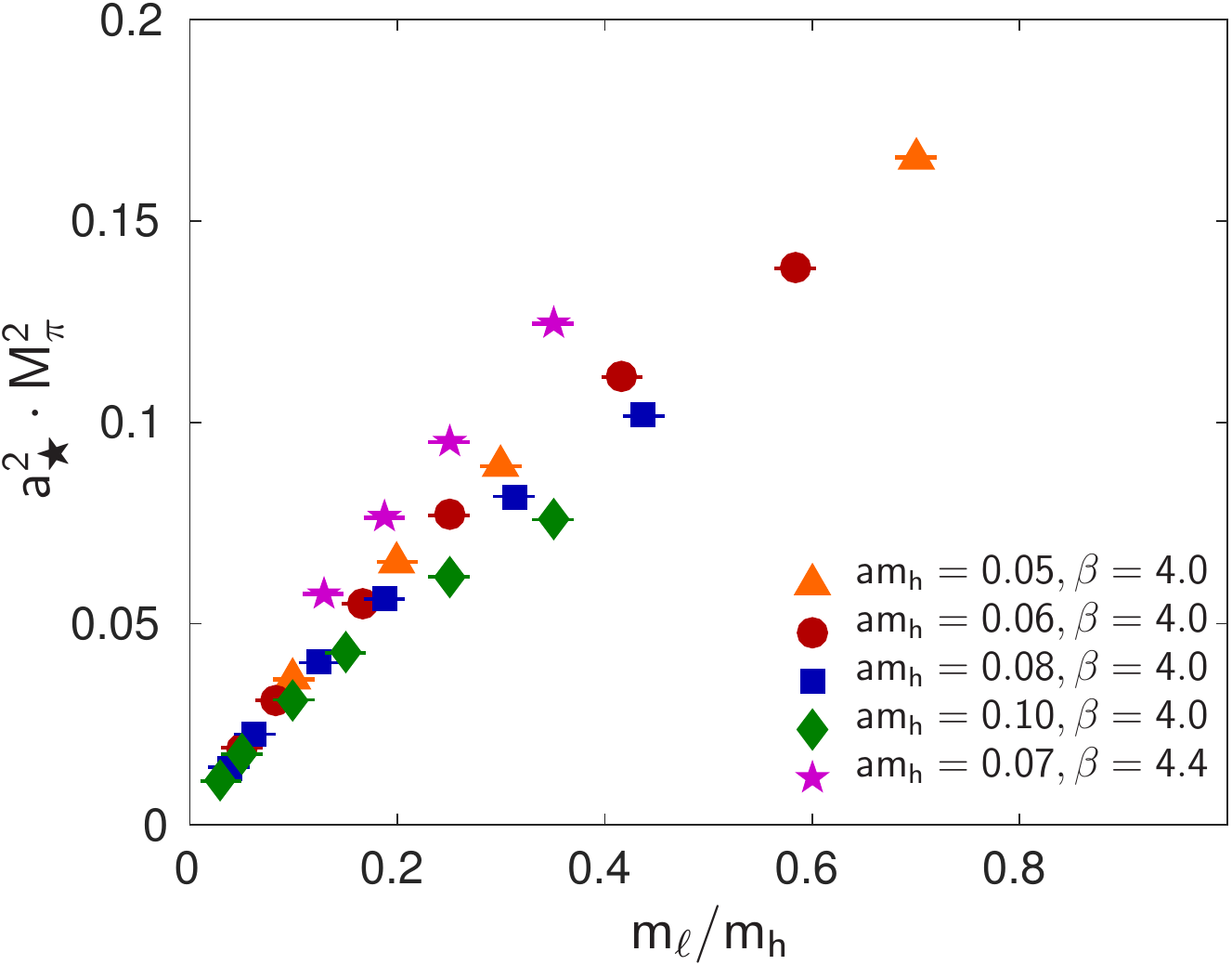}}
  \caption{Numerical evidence to show our system with four light and eight heavy flavors is chirally broken. Left: pseudoscalar decay constant $F_\pi$ vs.~the ratio of flavor masses $m_\ell/m_h$; right: squared mass of the pseudoscalar meson ($M_\pi$) vs.~$m_\ell/m_h$. Data are converted to the same lattice units, denoted by $a_\bigstar$, which matches the lattice spacing of the ensemble with ($\beta$, $m_\ell$, $m_h$) = (4.0, 0.003, 0.080).}
  \label{Fig.ChirallyBroken}
\end{figure}

First we demonstrate in Fig.~\ref{Fig.ChirallyBroken} that our system is indeed chirally broken: The left panel shows that the pseudoscalar decay constant $F_\pi$ takes a finite value in the chiral limit, whereas the right panel demonstrates that the squared mass of the pseudoscalar meson ($M_\pi$) is proportional to the mass of the light flavor $m_\ell$, if $m_\ell$ is sufficiently small. The left plot also shows very clearly, that simulations at different values for the heavy flavor mass $m_h$ or bare gauge coupling $\beta$ map out a unique curve only depending on $m_\ell/m_h$. This is an example of hyperscaling which arises due to the proximity of an IRFP.  Hyperscaling is also visible in $a_\bigstar^2 M_\pi^2$ (right plot) when considering $m_\ell/m_h \lesssim 0.2$. For larger values of $m_\ell/m_h$, corrections to scaling grow and are further enhanced by squaring $a_\bigstar M_\pi$. 

\begin{figure}[tb]
  \centering
  \parbox{0.086\textwidth}{\includegraphics[height=0.245\TXTheight]{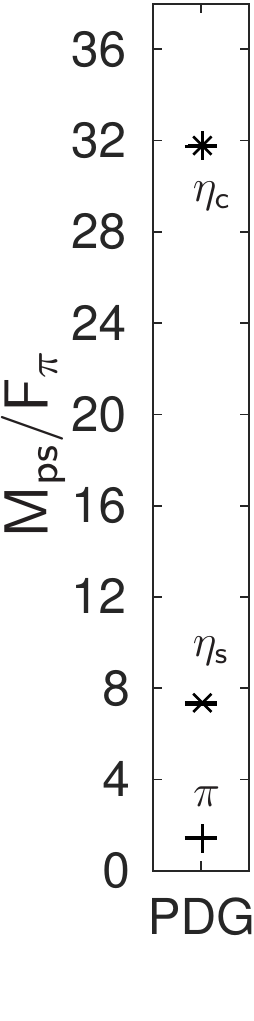}}
  \parbox{0.205\textwidth}{\includegraphics[height=0.245\TXTheight]{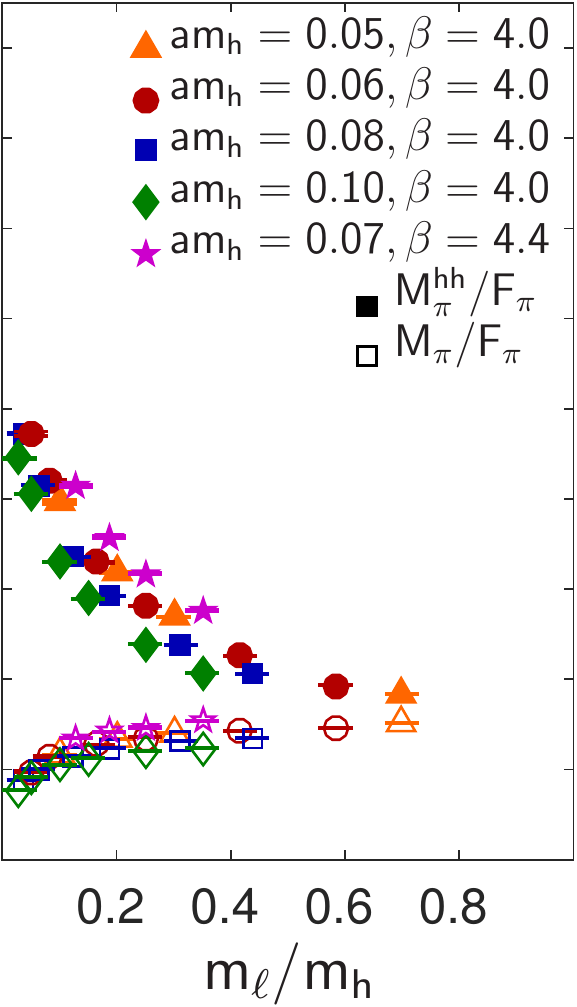}}
  \parbox{0.035\textwidth}{\includegraphics[height=0.245\TXTheight]{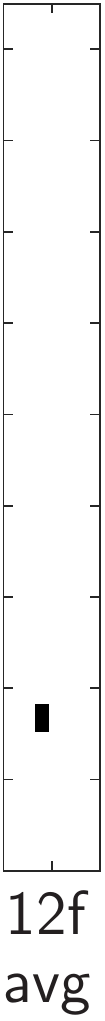}}
  \parbox{0.086\textwidth}{\includegraphics[height=0.245\TXTheight]{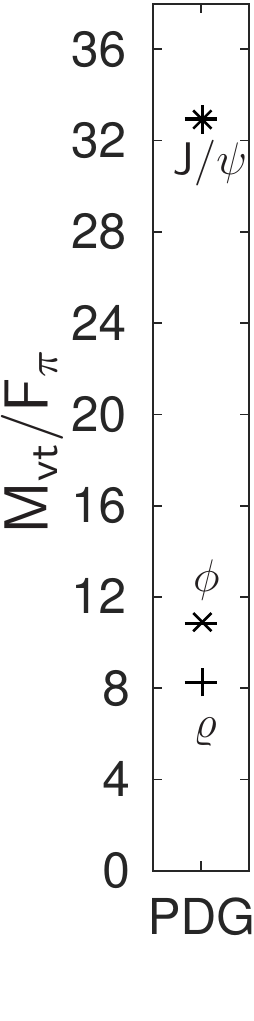}}  
  \parbox{0.205\textwidth}{\includegraphics[height=0.245\TXTheight]{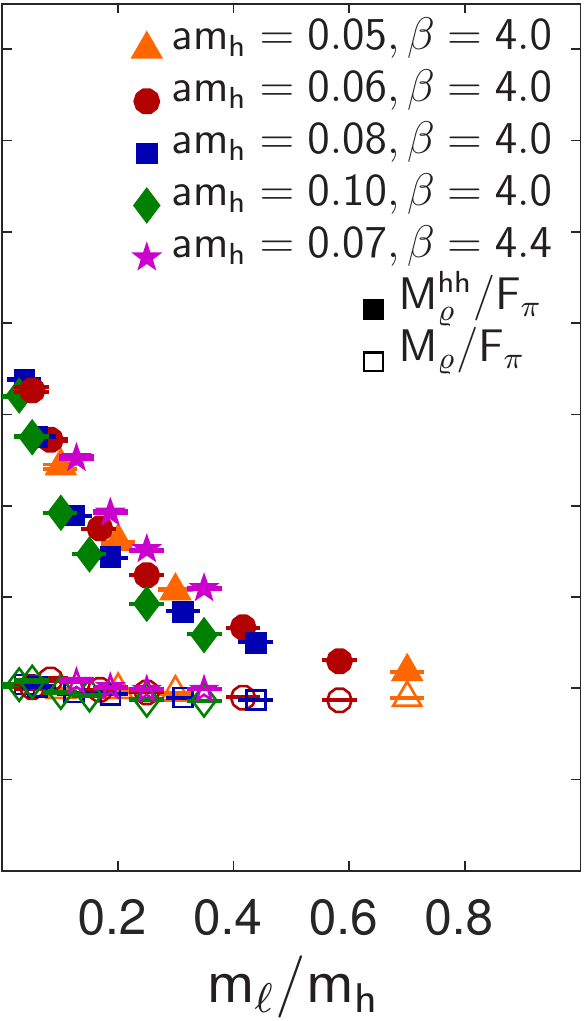}}
  \parbox{0.035\textwidth}{\includegraphics[height=0.245\TXTheight]{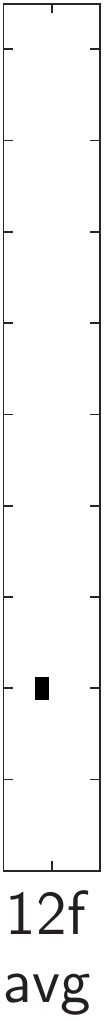}}
  \hfill
  \parbox{0.086\textwidth}{\vspace{-0.2mm}\includegraphics[height=0.244\TXTheight]{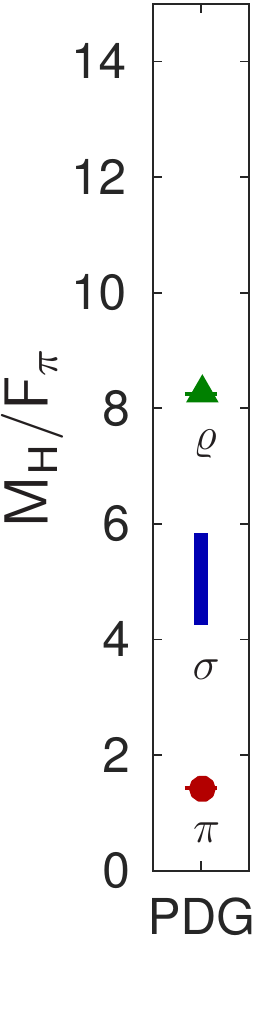}}
  \parbox{0.205\textwidth}{\includegraphics[height=0.245\TXTheight]{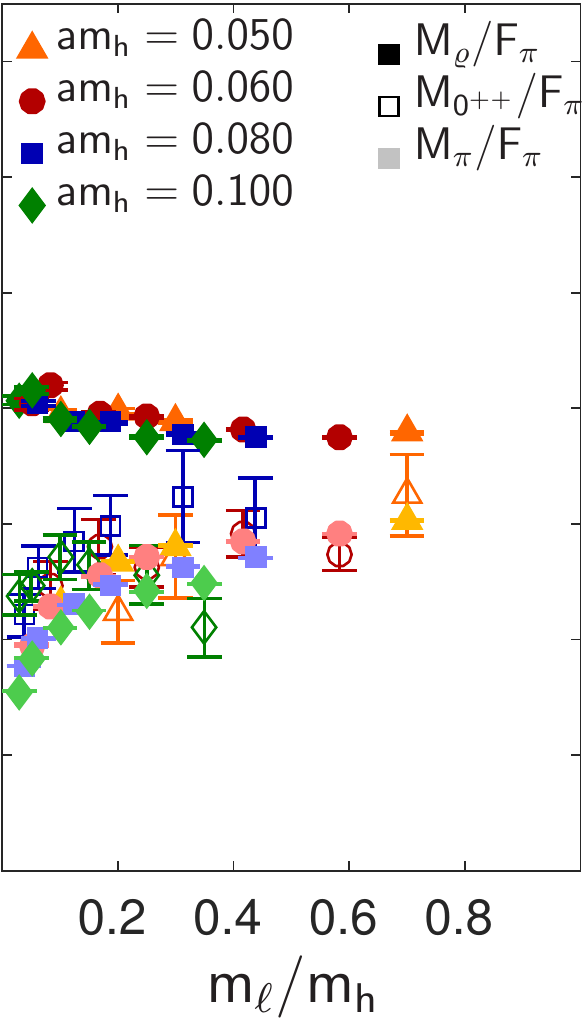}}
  \caption{The plots on the left show dimensionless ratios of meson masses in units of $F_\pi$ for pseudoscalar (ps) and vector (vt) states, respectively. The wide central panels show our data (with statistical errors only) as  function of $m_\ell/ m_h$. Different colors and symbols indicate the different $m_h$ and $\beta$ values, while filled (open) symbols denote states of the heavy-heavy (light-light) spectrum. The small panels to the right show averaged values for degenerate 12 flavors \cite{Aoki:2012eq,Fodor:2011tu,Cheng:2013xha,Aoki:2013zsa} and the panels on the left the corresponding PDG values \cite{Agashe:2014kda} and a pure $(s\bar s)$ pseudoscalar \cite{Dowdall:2013rya} for QCD divided by $F_\pi=94$ MeV. Due to the presence of an IRFP, the system shows hyperscaling and we observe independence of $m_h$, an unusual behavior in QCD standards. The plot on the right show the comparison of the  vector (filled), scalar (open), and pseudoscalar (shaded) states in units of $F_\pi$. PDG values for QCD \cite{Agashe:2014kda} are shown in the small panel, while our values for light-light states are shown in the wide panel. Within the reach of our simulations, the $0^{++}$ scalar is close to/degenerate with the pion.}
      \label{Fig.Spectrum}
\end{figure}

Next we focus at hyperscaling in hadron masses and how this constrains the spectrum. Focusing at the two plots on the left in Fig.~\ref{Fig.Spectrum}, we show in the central panels our data for the pseudoscalar and vector meson masses in units of the pseudoscalar decay constant $F_\pi$. The filled (open) symbols show states formed of light (heavy) flavors. Again our data obtained at different $\beta$ and $m_h$ values reveal hyperscaling in $m_\ell/m_h$ for all states. When $m_\ell/m_h \to 1$, light-light and heavy-heavy states get close to each other and approach the value obtained from mass-degenerate 12-flavor simulations (small panel to the right). On the left we show for comparison corresponding states listed by the PDG. While in QCD quarkonia are expected to scale proportional to the constituent quark mass, the proximity of an IRFP forces degeneracy when varying the mass of the heavy flavors. This makes the heavy-heavy spectrum rather distinct from QCD but only a few times heavier than the corresponding light-light states. In addition we also expect hyperscaling for the heavy-light states which energetically should lie between light-light and heavy-heavy states. Except for relating $F_\pi$ to the vacuum expectation value (vev) of the SM, there is no free parameter in our model  resulting in high predictivity.

Of special interest is the iso-singlet scalar ($0^{++}$) meson because in dilaton-like scenarios it is the candidate for the Higgs boson and numerical investigations of conformal or near-conformal theories have found that it is much lighter than the vector resonance,  often degenerate or even below the pseudoscalar \cite{Aoki:2013xza,Appelquist:2016viq,Aoki:2016wnc,Aoki:2012eq,Aoki:2013zsa,Kuti2016}. Determining the mass of the iso-singlet scalar is  challenging because the iso-singlet scalar has the same quantum numbers as the vacuum. It receives contributions from diagrams only connected by gluon lines. Thus its determination is much more noisy (for details see \cite{Weinberg:2014ega}). For a subset of our ensembles we have sufficient statistics to estimate the $0^{++}$ and show our results in the right most plot of Fig.~\ref{Fig.Spectrum}. We find a $0^{++}$ which is close/degenerate with the pseudoscalar and lighter than the vector. Unfortunately, additional simulation will be required to say with certainty where the $0^{++}$  ``peels-off'' from the pion and to predict its  chiral limit value.

\section{Conclusion}
We presented results from nonperturbative lattice field theory simulations to explore composite Higgs models exhibiting by construction a large separation of scales. Using numerically convenient staggered fermions, we study a model with four light and eight heavy flavors and demonstrated that this model is chirally broken but also shows hyperscaling in $m_\ell/m_h$. Due to hyperscaling the mass of heavy-heavy quarkonia is --- unlike in QCD --- not proportional to the constituent quark mass but depends only on $m_\ell/m_h$. Furthermore, this system has an iso-singlet scalar ($0^{++}$) state whose mass is much lighter than the vector and close to/degenerate with the pseudoscalar. 

To further investigate properties of mass-split systems, we are studying a system with four light and six heavy flavors using domain-wall fermions. Continuum-like symmetries of domain-wall fermions significantly simplify exploring the mechanisms to generate mass for the SM fermions.

\begin{acknowledgments}
The authors thank their colleagues in the LSD Collaboration for fruitful and inspiring discussions. 
Computations for this work were carried out in part on facilities of the USQCD Collaboration, which are funded by the Office of Science of the U.S.~Department of Energy, on computers at the MGHPCC, in part funded by the National Science Foundation (award OCI-1229059), and on computers allocated under the NSF Xsede program to the project TG-PHY120002. 
We thank Boston University, Fermilab, the NSF and the U.S.~DOE for providing the facilities essential for the completion of this work.  A.H. acknowledges support by DOE grant 
DE-SC0010005 and C.R. by DOE grant DE-SC0015845.   This project has received funding from the European Union's Horizon 2020 research and innovation programme under the Marie Sk{\l}odowska-Curie grant agreement No 659322. 
\end{acknowledgments}

{\small
  \bibliography{../General/BSM}
  \bibliographystyle{apsrev4-1}
}

%\begin{thebibliography}{99}
%\bibitem{...}
%....
%\end{thebibliography}

\end{document}